\documentclass[useAMS, usenatbib]{mn2e}

\usepackage{graphicx, bm, amssymb, xcolor}
\usepackage{verbatim}
\usepackage{appendix}

\usepackage{threeparttable}
\usepackage[backref,breaklinks,colorlinks,citecolor=blue]{hyperref}
\usepackage[all]{hypcap}
\usepackage{times}

\newcommand{\noopsort}[1]{}

\usepackage{newtxtext,newtxmath}
\usepackage[T1]{fontenc}
\usepackage{ae,aecompl}

\usepackage{graphicx}	% Including figure files
\usepackage{amssymb}	% Extra maths symbols

\usepackage{epsfig}
\usepackage{subfigure}
\def \aj{Astronom.~J.}

\def \aap{A\&A }
\def \apjl{ApJ}
\def \apjs{ApJS}
\def \apj{ApJ}

\def \mnras{MNRAS}

\def\zsun{{\,Z_\odot}}

\newcommand{\physrep}{Phys.~Rep.}
\newcommand{\araa}{ARA\&A}

\newcommand{\fE}{f_{\rm E,*}}

\newcommand{\fEc}{f_{\rm E,crit}}
\newcommand{\fEcr}{f_{\rm E,crit,rt}}
\newcommand{\fEcc}{f_{\rm E,crit,c}}

\newcommand{\msun}{M_{\odot}}
\newcommand{\lsun}{L_{\odot}}

\newcommand{\fEavg}{\left\langle  f_{\rm E}\right\rangle}

\newcommand{\Sigoc}{\Sigma_{0,\rm{crit}}}

\def\alt{\raise0.3ex\hbox{$\;<$\kern-0.75em\raise-1.1ex\hbox{$\sim\;$}}}
\def\agt{\raise0.3ex\hbox{$\;>$\kern-0.75em\raise-1.1ex\hbox{$\sim\;$}}}

\newcommand{\lsim}{\,\rlap{\raise 0.35ex\hbox{$<$}}{\lower 0.7ex\hbox{$\sim$}}\,}
\newcommand{\gsim}{\,\rlap{\raise 0.35ex\hbox{$>$}}{\lower 0.7ex\hbox{$\sim$}}\,}

\defcitealias{Crocker2018}{CKTC18}
\defcitealias{Fall2010}{FKM10}

\title[Radiation Pressure in Compact Stellar Systems]{Radiation Pressure Limits on the Star Formation Efficiency and Surface Density of Compact Stellar Systems}

\author[R. M.~Crocker et al.]{Roland M.~Crocker,$^{1}$\thanks{E-mail: rcrocker@fastmail.fm (RMC)}
Mark R. Krumholz,$^{1,2}$
Todd A. Thompson,$^{3}$
\newauthor 
Holger Baumgardt,$^{4}$
and Dougal Mackey$^{1,2}$ \\
\\
$^{1}$Research School of Astronomy and Astrophysics, Australian National University, Canberra 2611, A.C.T., Australia\\
$^{2}$Centre of Excellence for Astronomy in Three Dimensions (ASTRO-3D), Australia\\
$^{3}$Department of Astronomy and Center for Cosmology \& Astro-Particle Physics, The Ohio State University, Columbus, Ohio 43210, U.S.A\\
$^{4}$School of Mathematics and Physics, University of Queensland, Brisbane 4072, Australia\\
}

\date{Accepted XXX. Received YYY; in original form ZZZ}

\pubyear{2018}

\begin{document}
\label{firstpage}
\pagerange{\pageref{firstpage}--\pageref{lastpage}}
\maketitle

\begin{abstract}
The large columns of dusty gas enshrouding and fuelling star-formation in young, massive stellar clusters
may render such systems optically thick to radiation well into the infrared.
This raises the prospect that both ``direct'' radiation pressure produced by absorption of photons leaving stellar surfaces and ``indirect'' radiation pressure from photons absorbed and then re-emitted by dust grains may be important sources of feedback in such systems.
Here we evaluate this possibility by deriving the conditions under which a spheroidal, self-gravitating, mixed gas-star cloud
can avoid catastrophic disruption by the combined effects of direct and indirect radiation pressure. 
We show that radiation pressure sets a maximum star cluster formation efficiency of $\epsilon_{\rm max} \sim 0.9$ at a 
(very large)
gas surface density of $\sim 10^5 \msun$ pc$^{-2} (\zsun/Z)
\simeq 20$ g cm$^{-2} (\zsun/Z)$,
but that gas clouds above this limit undergo significant radiation-driven expansion during star formation,
leading to a maximum stellar surface density very near this value for all star clusters.
Data on the central surface mass density of  compact stellar systems, while sparse and partly confused by dynamical effects, are broadly consistent with the existence of a metallicity-dependent upper-limit  comparable to this value.
Our results imply that this limit may preclude the formation of the progenitors of intermediate-mass black holes  for systems with $Z \gsim 0.2 \zsun$.
\end{abstract}

\begin{keywords}
hydrodynamics -- instabilities-- ISM: jets and outflows -- radiative transfer -- galaxies: ISM -- galaxies: star clusters
\end{keywords}

%%%%%%%%%%%%%%%%%%%%%%%%%%%%%%%%%%%%%%%%%%%%%%%%%%

%%%%%%%%%%%%%%%%% BODY OF PAPER %%%%%%%%%%%%%%%%%%

\section{Introduction}
\label{sec:intro}

It has been appreciated for some time that the direct radiation flux from young stars 
may be sufficiently intense to drive gas out of isolated protoclusters experiencing intense star formation.
As such, direct radiation pressure -- i.e., the momentum flux imparted by starlight to gas mediated via the photons' initial absorption and scattering by the dust borne by the gas
-- is an important agent of ``feedback'' in such systems, 
a view supported by both observational  \citep{Scoville2001,Lopez2011,Lopez2014} and theoretical
\citep{Krumholz2009a,Fall2010,Murray2010b,Murray2011,Skinner2015,Thompson2016} perspectives.

In this article -- which leaves off from developments in our previous paper on star-forming discs \citep[][hereafter \citetalias{Crocker2018}]{Crocker2018} which, in turn, follows \citet{Krumholz2012,Krumholz2013}  -- we show that {\it indirect} radiation pressure \citep[i.e., radiation pressure due to dust-reprocessed photons rather than direct starlight photons; cf.][]{Murray2010b} will also, in many cases, have an important dynamical effect in nascent star clusters.
Indirect radiation pressure effects arise because the molecular gas, from which stars form, bears dust that reradiates absorbed UV and optical light at infrared (IR) wavelengths.
This same dust may -- if it presents a sufficiently large column -- subsequently scatter or absorb and reradiate the starlight down-scattered into the IR.
Indeed, at the large gas and, consequently, dust columns concomitant to the intense star formation surface densities encountered in local ultra-luminous infrared galaxies
(such as Arp 220) or in sub-mm galaxies at higher redshifts, a star-forming environment may be optically thick to photons with wavelengths as long 
$\sim 100 \ \mu$m\footnote{Fits to such galaxies' spectral energy distributions imply gas columns of $\sim 0.01 - 1$ g cm$^{-2}$ \citep{Chakrabarti2008}, corresponding to optical depths of $\sim 10 - 100$ at 20 $\mu$m and $\sim 1 - 10$ at 100 $\mu$m.}.
Most pertinent here, similar or even larger dust columns and consequent optical depths are also encountered in the densest star-forming clouds found in the Milky Way
and other galaxies in the local Universe. For example, \citet{Clarkson2012} find that the stellar surface density in the central $0.4$ pc of the Arches cluster near the Galactic Centre is $\approx 4$ g cm$^{-2}$, and this is only a lower limit on the initial gas surface density, corresponding to optical depths of several at 100 $\mu$m. The young or still-forming super star clusters in M82 \citep{McCrady2007} and NGC 253 \citep{Leroy2018} have gas columns exceeding $10$ g cm$^{-2}$, so at 100 $\mu$m the optical depth is $\sim 10$.

The existence of large optical depths at IR wavelengths in such systems raises the interesting possibility that, as the downshifted photons bang around inside the gas column, 
much more momentum can be extracted from them than is possible in the single-scattering limit pertinent to direct radiation pressure.
Indeed, in the ``strong-trapping'' limit the momentum per unit time extracted from starlight is amplified from $L/c$ to $\sim \tau L/c$, where $\tau$ is the optical depth\footnote{Ultimately only bound, in principle, by energy conservation to $\lsim (c/v) L$ where $v$ is a characteristic velocity of the outflowing gas \citep[e.g.,][]{Lamers1999}.}. 
While such a trapped radiation field could in principle eject gas from galaxies \citep{Murray2011}, whether it does so in practice is a separate question. Both analytic calculations and simulations suggest that dust-reprocessed radiation pressure is a threshold effect: for a given column of gas and stars, there exists a critical maximum radiation flux that can be forced through the column while the gas remains in a stable hydrostatic equilibrium. As long as such an equilibrium exists, the gas plus radiation pressure forces self-adjust to be in balance with the gravitational force, and IR radiation pressure has little effect on the dynamics. Once the flux exceeds the critical value, however, the gas column becomes unstable and turbulent, and simulations show that the instability saturates into a state where the mass-averaged radiation force exceeds (though, as a result of radiation Rayleigh-Taylor instability, only slightly) the mass-averaged gravitational force, rendering the entire gas column super-Eddington and thus liable to ejection on a dynamical timescale \citep[e.g.,][]{Davis2014, Tsang2015, Zhang2017}. Thus the question of whether indirect radiation pressure is important for star formation reduces to the question of whether star forming systems have combinations of gas column and radiative flux that place them into the stable and sub-Eddington regime or the unstable and super-Eddington one.

In \citetalias{Crocker2018} we showed that the radiative flux in most star-forming discs, averaged over the entire star-forming disc, fall into the former rather than the latter category. 
Consequently, for the vast majority even of starbursting galaxies, radiation pressure cannot be an important regulator of star formation or driver of winds at galactic scales.
On the other hand, %we also found in \citetalias{Crocker2018} that 
as first shown by \citet{Thompson2005} and later confirmed in \citetalias{Crocker2018},
there is a very suggestive coincidence \citep[cf.][]{Andrews2011} between the upper boundary of the occupied region of the Kennicutt-Schmidt (KS) parameter space (of star formation surface density $\dot{\Sigma}_{*}$ vs.~gas surface density $\Sigma_{\rm gas}$) and the critical value of the star formation surface density
($\dot{\Sigma}_{\rm \star,crit} \sim 10^3 \msun$ pc$^{-2}$ Myr$^{-1}$, corresponding to a critical radiative flux of $F_{\rm \star,crit} \sim 10^{13} \lsun$ kpc$^{-2}$) where indirect radiation pressure effects preclude hydrostatic equilibrium.
This strongly suggests that, while indirect radiation pressure is not an important agent of feedback {\it on  global scales} for most systems, it nevertheless circumscribes the parameter space along the locus of the KS relation that may be occupied by real systems.
Moreover, even if a star forming system is globally sub-Eddington, the fact that star formation is highly clumpy can render individual star-forming sub-regions 
(i.e.,
nascent star clusters and super star clusters forming out of individual giant molecular clouds)
super-Eddington, driving outflows away from these sub-regions \citep{Murray2011}.
Numerical exploration of this question in the context of the most massive star clusters, where the effect is expected to be most important, has been highly limited by computational costs. \citet{Skinner2015} and \citet{Tsang2018} both simulate indirect radiation pressure effects in a forming star cluster, and find them to be quite modest, but they survey a very small part of parameter space, and one where, as we show below, indirect radiation pressure effects are not expected to be significant. Moreover, these authors adopt a constant opacity to indirect radiation pressure, an assumption that we showed in \citetalias{Crocker2018} leads to fundamental changes in the dynamics of the problem. Motivated by the need to explore this problem over a wider range of parameters than simulations currently permit,
here we direct our attention to the importance of indirect radiation and to its 
effects on 
local scales within regions experiencing intense star formation, e.g., individual molecular clumps that are collapsing to form stellar proto-clusters or, in principle, larger stellar spheroids. 
Furthermore,  we simultaneously incorporate {\it direct} radiation pressure effects in our model following the treatment of \citet[hereafter \citetalias{Fall2010}]{Fall2010}.

One might suspect that a threshold feedback mechanism such as indirect radiation pressure might be important for star-forming subregions because \citet{Hopkins2010} have compiled observations
that support the existence of a universal, maximum {\it central}\footnote{That is, {\it not} the effective or mean surface density $M_{\star}/(\pi R_e^2)$ but the limiting surface mass density as $r \to 0$.} surface mass density $\Sigma_{\rm max} \sim 10^5 \msun/$pc$^{-2} \simeq 20$ g cm$^{-2}$ for spheroidal stellar systems across a  span of $\sim$7 orders of magnitude in total stellar mass $M_{\star}$ and $\sim$5 in effective radius $R_e$.
These dense stellar systems include globular clusters in the Milky Way and nearby galaxies, massive star clusters in nearby starbursts, nuclear star clusters in dwarf spheroidals and late-type discs, ultra-compact dwarfs, and galaxy spheroids spanning the range from low-mass ``cusp'' bulges and ellipticals to massive ``core'' ellipticals; these systems are all baryon dominated and likely formed in rapid, dissipational events. 
\citet{Hopkins2010} have already weighed a number of theoretical scenarios that might explain the existence of a universal $\Sigma_{\rm max}$
including that it is an effect of infrared radiation pressure mediated by dust.
While seen as particularly promising by \citet{Hopkins2010}, these authors did note that an explanation invoking indirect radiation pressure faces a number of challenges and further, recent work involving some of the original authors \citep{Grudic2018a,Grudic2018b} seems to cast further doubt on this general scenario.
The most significant challenge facing the scenario is presented by the differing metallicities of these systems and, therefore, the well-motivated expectation that their dust-to-gas ratio -- and therefore infrared opacity -- should vary.
In particular, one would expect {\it ceteris paribus} that systems that form their stars at quite high redshift and which are observed to have low stellar metallicities today, would have higher maximum surface densities.
Other challenges include that the mechanism would  seem not to work if the stellar content of the system were built up in a number of star-formation events and that, again for systems formed at high redshift, there has been a lot of time for dynamical relaxation to 
diminish the central surface mass density. 
We revisit the question of whether the observed maximum stellar surface density can in fact be explained by radiation pressure effects in this work.

In brief, this paper extends previous work in three ways: i) we determine the stability curve for star forming, dusty clouds subject to radiation pressure 
deriving from the nascent stars
adopting a realistic temperature-dependent opacity $\kappa \propto T^2$ (rather than assuming a constant value);
ii) we account simultaneously for direct and indirect radiation pressure effects;
and iii) we follow the evolution of star clusters (forming out of giant molecular clouds) accounting for radiation pressure
in our determination of the final state surface mass density as a function of initial state surface mass density.
The remainder of this paper is as follows. In \autoref{sec:model} we present a theoretical calculation for the process of star formation limited by both direct and dust-reprocessed radiation. In \autoref{sec:discussion} we compare this theoretical calculation to observations of star clusters, and discuss further implications of our findings. We summarise and conclude in \autoref{sec:summary}.

\subsection{A word on notation and symbols}

To forestall any potential confusion, below we label (dimensionful and dimensionless) reference values of some general quantity $x$
 with a subscript asterisk: $x_*$.
Thus, for instance, $T_*$ is a reference (dimensionful) temperature and $\tau_*$ is a (dimensionless) reference optical depth (both defined below).
Quantities connected to stars or star formation are labelled with a subscript ``$_\star$".
Thus, for instance, $M_\star$ is the total stellar mass of the system and $\dot{\Sigma}_{\rm \star,crit}$ is the critical star formation surface density (defined below).

%%%%%%%%%%%%%%%%%%%%%%%%%%%%%%%%%%%%%%%%%%%%%%%

\section{Radiation pressure effects in star-forming clouds}
\label{sec:model}

Our goal in this section is to derive a theoretical model for the joint effects of direct and indirect radiation pressure feedback on a star-forming cloud. We begin in \autoref{ssec:CKTC18} by considering the effects of indirect radiation pressure, summarising the relevant results from \citetalias{Crocker2018} and adapting them from planar to spherical geometry. We then derive evolutionary tracks for star-forming clouds subject to indirect radiation pressure effects in \autoref{ssec:tracks_indirect}. We extend this analysis to include the effects of direct radiation pressure, based on the treatment of \citetalias{Fall2010}, in \autoref{ssec:tracks_full}.

\subsection{Indirect radiation pressure}
\label{ssec:CKTC18}

\subsubsection{Recapitulation of \citetalias{Crocker2018}}

The fundamental calculation in \citetalias{Crocker2018} is as follows: for a specified planar column of self-gravitating, dusty gas and stars with a specified turbulent velocity dispersion and internal radiation source, we search for a density and temperature profile that allows the system to be in simultaneous hydrostatic and radiative equilibrium. We then show that such equilibria exist if the radiation flux is below a critical value, and that radiation fluxes above this value destabilise the gas, causing it to become turbulent and driving much of it off in a wind. The system we consider in \citetalias{Crocker2018} is, in the approximation that the local radiation spectrum is always well described by a Planck function and that dust and gas are dynamically-well coupled,
fully characterised by its surface densities of gas $\Sigma_{\rm gas}$ and stars $\Sigma_{\star}$, by the radiation flux per unit area provided by the stars\footnote{Of course, very similar considerations  around, e.g., the stability (or not) of the gas column (see below), apply in the case that
the radiative flux is dominantly supplied by an active galactic nucleus rather than star formation, but the physical context under consideration here is that of an isolated gas cloud collapsing to form a stellar cluster or super cluster; we therefore limit ourselves to consideration of the case that the radiative flux is supplied by stars.} $F_\star$, and by the temperature-dependent Rosseland mean opacity of the dusty gas $\kappa_R(T)$; for $T \lesssim 150-200$ K, the opacity is well-approximated by a powerlaw $\kappa_R\propto T^2$ \citep{semenov03a}\footnote{We show in \autoref{sec:T0appendix} that the systems we consider in this paper satisfy this constraint.}. From these dimensional quantities, \citetalias{Crocker2018} (partially following \citealt{Krumholz2012, Krumholz2013}) show that one can define three dimensionless parameters that, together with the functional form of $\kappa_R(T)$, determine the stability of the system:
\begin{eqnarray}
f_{\rm gas} & = & \frac{\Sigma_{\rm gas}}{\Sigma_{\rm gas}+\Sigma_{\star}} \\
\tau_* & = & \frac{1}{2}\Sigma_{\rm gas} \kappa_{R,*} \\
\fE & = & \frac{\kappa_{R,*} F_\star}{g_* c}
\label{eqn:fE}
\end{eqnarray}
where $T_* = (F_\star/ca)^{1/4}$ is the reference temperature at the top of the column, $\kappa_{R,*} = \kappa_R(T_*)$ is the opacity evaluated at this temperature, and $g_* = 2 \pi G (\Sigma_{\rm gas} + \Sigma_{\star})$ is the gravitational acceleration at the top of the column. Intuitively, the quantities $f_{\rm gas}$, $\tau_*$, and $\fE$ are the gas fraction, the effective optical depth, and the effective Eddington ratio, with the latter two quantities computed using the conditions that prevail at the top of the atmosphere. We emphasise that $\fE$ is distinct from $\fEavg$, which we define as the mass-weighted mean Eddington ratio throughout the gas column; the former is computed using conditions at the top of the gas column, while the latter is a mass-weighted mean over the gas column, and thus depends on the configuration of mass and radiation flux.

While $\fE$ and $\fEavg$ are physically distinct,~\citetalias{Crocker2018} show that the former determines the latter.
For any specified gas fraction and optical depth, and for an opacity law $\kappa_R \propto T^2$ as expected at the temperatures relevant for star-forming molecular gas, there exists a maximum value of the Eddington ratio $\fE = f_{\rm E,crit}(\tau_*, f_{\rm gas})$, and thus a maximum radiation flux $F_\star$, at which such a gas column can be in hydrostatic balance.\footnote{To be precise, \citetalias{Crocker2018}'s calculation of $f_{\rm E,crit}(\tau_*, f_{\rm gas})$ depends on the assumption one makes about the efficiency of convective transport in regions where the column is unstable to convection, but where conventional convection is unable to carry a significant heat flux because radiation rather than gas dominates the enthalpy budget. \citetalias{Crocker2018} derive values of $f_{\rm E,crit}(\tau_*, f_{\rm gas})$ for the two limiting cases of efficient convection (i.e., convection is able to carry enough heat to fully flatten the entropy gradient) and zero convective heat flux. 
(These two cases are labelled $\fEcc$ and $\fEcr$,  where c (rt) means convection (radiative transfer) governs the temperature profile.)
The two differ substantially only at gas fractions above $\approx 50\%$. 
For our purposes in this paper we will mostly use the efficient convection limit, since simulations suggest reality lies closer to that case, only referencing the inefficient convection limit on occasion to point out the overall similarity of the results it produces. See \citetalias{Crocker2018} for further discussion and cf. \autoref{sec:RTappendix}.
\label{footnote:CandRT}} 
If the system satisfies $\fE < f_{\rm E,crit}$, then the mass arranges itself into a vertical density profile for which  $\fE < \fEavg <1$, and the system is both stable and sub-Eddington. 
(We use the self-consistently calculated density -- and consequent temperature -- profiles from \citetalias{Crocker2018} in the modelling presented in this paper.)
If $\fE > f_{\rm E,crit}$, on the other hand, no equilibrium matter profile is possible and, in fact, the system is unstable to the development of radiation Rayleigh-Taylor instability. 
Numerical simulations then show that in the saturated state of the instability $\fEavg$ asymptotes to a value close to but just above unity, so that mass is slowly driven upward out of the gravity well by the radiation force.
In physical units, for Milky Way dust opacity, and in a planar geometry, the condition that $\fE > f_{\rm E,crit}$ and thus that gas be ejected is approximately equivalent to the condition that the radiation flux $F_\star \gtrsim 10^{13} \lsun $ kpc$^{-2}$ \citep[a  scale previously identified in][]{Thompson2005} corresponding to a maximal star formation
surface density $\sim 10^{3} \msun $ pc$^{-2}$  Myr$^{-1}$.

\subsubsection{From planar to spherical geometry: limiting cases}

A gas cloud that is collapsing to form a star cluster will differ from the case considered in \citetalias{Crocker2018} in that its geometry is primarily 3D rather than 2D. We must therefore extend the \citetalias{Crocker2018} calculation from approximately planar to approximately spherical geometry. 
The column density of the cloud remains a critical variable, even in spherical geometry, because this is what will determine the run of temperature through the cloud, as first shown by \citet{Chakrabarti2005, Chakrabarti2008}, and the temperature determines the opacity and thus the radiation pressure force. A second issue to consider in spherical geometry is that the column density will not be uniform due to the clumpiness of the gas. However, we show in \autoref{app:clumpy} that this has relatively modest effects, and induces uncertainties in our final results only at the factor of few level. However, the change from planar to spherical geometry nonetheless
has an important effect on the stability of the system against indirect radiation pressure. 
While in spherical geometry the gravitational force and radiation {\it flux} both decline as $1/r^2$ (for $r$ large enough that most of the mass is enclosed),  the {\it force} exerted by indirect radiation pressure on gas at the surface of a cloud falls faster than $1/r^2$ because, as the cloud's radius expands, the dust temperature and opacity decline (again assuming that $T \lesssim 150-200$ K, so that we have $\kappa_R \propto T^2$). In terms of our dimensionless parameters, the key distinguishing characteristic of spherical geometry is that $\kappa_{R,*}$ is a decreasing function of cloud radius, so that $\fE$ falls with radius as well.
This differs from the planar case, for which there is no flux divergence and thus the opacity goes to a fixed, non-zero value as $z \to \infty$, rather than dropping to zero, so that $\fE$ is constant.
It also differs from the case of the direct radiation force treated in \citetalias{Fall2010} and that we shall revisit below, where the opacity is determined by the colour temperature of the stellar sources and thus is independent of distance.  

Because the indirect radiation force in spherical geometry drops with radius faster than the gravitational force, there is no maximum flux beyond which it becomes impossible for a gas column to remain hydrostatic. 
At sufficient distance, the IR radiation force always drops to zero (in terms of our dimensionless parameters, $\fE \rightarrow 0$), and thus for a sufficiently extended spherical configuration a solution that looks like a pressure-supported atmosphere always exists. (Recall that this statement applies only to the IR force; we will address the question of the direct radiation force below.)
In light of this discussion, we can ask what happens to a star-forming cloud that, as a result of a rising luminosity as gas transforms into stars, begins to violate the planar stability condition derived by \citetalias{Crocker2018} at its surface (i.e., $\fE > f_{\rm E,crit}$ using the radiation flux and surface density evaluated at the cloud radius $r$). If $\fE > f_{\rm E,crit}$ in planar geometry, then the mass-weighted mean Eddington ratio $\fEavg > 1$, and there is no way that the system can return to a sub-Eddington state as long as the flux remains fixed. For spherical geometry, on the other hand, a system that has $\fEavg > 1$ at its starting radius will at some point expand enough to have $\fEavg < 1$, because $\fE$ will drop below $f_{\rm E,crit}$. We can imagine two limiting outcomes for what will happen in this case:

\vspace{0.2cm}
\noindent
{\bf Type A (disruption):} any time a spherical cloud finds itself in a configuration with $\fE > f_{\rm E,crit}$, it expands violently, leaving it vulnerable to rapid disruption by something else: hydrodynamic stripping by background flows, loss of mass from supernovae, stellar winds,  or direct radiation pressure, or by tidal stripping. Under this assumption a cloud that becomes super-Eddington is assumed to disrupt almost immediately. 

\vspace{0.2cm}
\noindent
{\bf Type B (quasi-equilibrium expansion):} a cloud that finds itself with $\fE > f_{\rm E,crit}$ undergoes quasi-equilibrium expansion as needed in order to remain marginally stable, $\fE = f_{\rm E,crit}(\tau_*, f_{\rm gas})$, at its instantaneous gas fraction and luminosity. It will continue to form stars until it is disrupted by direct radiation pressure or some other mechanism.

\vspace{0.2cm}

While neither of these two limiting cases for cloud evolution is likely to to exactly describe the messy, non-linear reality of cloud expansion under radiation pressure,
they
 will bracket reality.
 Moreover, as we show below, expressed in terms of an efficiency defined as the ratio of final to initial surface density, they lead to rather similar results.
We spend a few sentences here, however, discussing which scenario is likely to be closer to the truth.
This basically comes down to a question of relative timescales.
The timescale for the surface density to change is a dynamical time. 
However, the timescale for the temperature and thus opacity to change is significantly smaller than that, because the timescale to reach thermal equilibrium is the radiation diffusion time, which is the size of the system divided by the characteristic diffusion speed of $c /\tau$. 
Thus the thermal equilibration time is $\tau r / c$, which is smaller than the crossing time by a factor of $c / (\tau v) \gg 1$. 
Thus to good approximation the opacity that the radiation field sees adjusts instantaneously (with respect to the dynamical time).

The remaining question is whether, in a cloud that finds itself at a luminosity such that it exceeds the stability limit at its current radius, gas is accelerated rapidly enough to exceed the escape speed before it has time to move far enough out in radius for the spherical divergence of the radiation field to reduce the force back below the Eddington limit.
For indirect radiation pressure mediated by infrared radiation and powered by star formation, the answer to this is likely `no' and for the following reason: suppose we have a cloud with initial radius $r$ and surface gravitational acceleration $g$, which finds itself slightly super-Eddington because ongoing star formation makes the radiation flux passing through it large enough that $\fE > f_{\rm E,crit}$.
It will only remain super-Eddington until the material is able to expand significantly, so the condition for disruption is, roughly, that the gas be accelerated to above the escape speed before the radius changes by a factor of order unity. 
Suppose further that the net outward acceleration of the material is $a_{\rm net}$, which we can write in terms of an effective Eddington ratio for the accelerating gas as $a_{\rm net} \simeq (\fEavg - 1) g.$
Thus by the time the gas propagates from $r$ to $2 r$, its velocity will be $v \simeq \sqrt{2 r (\fEavg - 1) g}.$
The escape speed from radius $r$  is $\sqrt{2 g r}$, and from $2 r$ it is $\sqrt{g r}$. 
Thus the condition that $v > v_{\rm esc}$ is satisfied only if
$\sqrt{2 (\fEavg - 1)} > 1$.
However, numerical simulations of super-Eddington dusty atmospheres show that, as a result of RRT instability, $\fEavg$ is very close to unity even when gas is unstable and being ejected \citep{Krumholz2012,Krumholz2013,Davis2014,Tsang2015,Zhang2017}, so that $\fEavg - 1 \ll 1$. 
This in turn means that the IR radiation will not be sufficient to raise gas to the escape speed over a distance of radius $r$.\footnote{Note that this argument applies only if changes in the radiative flux occur on timescales that are not fast compared to the timescale over which the gas can dynamically readjust so as to maintain
$\fEavg \sim 1$. This condition is satisfied for systems where the power source is star formation, since the radiation flux provided by stars cannot increase much faster than the dynamical time of the system. However, it might not be satisfied for active galactic nuclei, where changes in the accretion behaviour of the small inner accretion disc can drive luminosity increases on timescales much shorter than the dynamical time of the vastly larger dusty torus.
Also note the  situation here is quite different to the direct radiation scenario explored in \citet{Thompson2016}. 
In that case, radiation pressure can drive out significant amounts of material because a non-negligible fraction of material (in low column regions) finds itself super-Eddington by a factor $\gsim 2$; such does not occur in the simulations of indirect radiation pressure. 
There might be some tail of fast-moving material that gets accelerated to escape speed, but the bulk certainly is not.}
Thus, while we certainly expect some gas to be driven out of a system as it forms stars, 
the effect of indirect radiation pressure in dusty, star-forming (spherical) clouds can be largely characterised as causing a quasi-equilibrium expansion to lower surface mass density (i.e., the `Type B' scenario).
Of course, this picture is complicated if there are other sources of feedback acting but, as we show below, even accounting for direct radiation pressure effects, this general picture does not change too much.

\subsection{Evolutionary tracks with indirect radiation pressure}
\label{ssec:tracks_indirect}

With the preceding discussion in mind, we now adapt our previous work  to the case of spherical, self-gravitating and  star-forming molecular clouds.
With $R$ the cloud radius,
the  total (cross-sectional) surface mass density is
\begin{equation}
\Sigma_{\rm tot} = \Sigma_{\rm *} + \Sigma_{\rm gas} = \frac{M_{\rm tot}}{\pi R^2} = \frac{M_{\star}}{(1 - f_{\rm gas}) \  \pi R^2}
\end{equation}
where, in the last equality, we utilise the gas (mass) fraction $f_{\rm gas}$.
The surface gravity is
\begin{equation}
g_* = \frac{G M_{\rm tot}}{R^2} \, .
\end{equation}
The luminosity of a simple stellar population of mass $M_{\star}$ and age $t$ that fully samples the IMF can be approximated by $L_\star \simeq \Psi M_{\star}/{\rm max}(1,t/t_{\rm cr})$ \citep{Krumholz2010, Crocker2018}, where $\Psi$ is the light to mass ratio of a zero age stellar population, and $t_{\rm cr}$ is the time required for the population to reach statistical equilibrium between an increase in luminosity from new stars forming and a decrease from older stars going out. 
For a Chabrier IMF, $\Psi = \Psi_0 \approx 2200$ erg s$^{-1}$ g$^{-1} \approx 1100$ $L_\odot$ $M_\odot^{-1}$ and $t_{\rm cr} = t_{\rm cr,0} \approx 6.9$ Myr.\footnote{Formally, $t_{\rm cr} \equiv \Phi/\Psi$, where $\Phi = 4.1\times 10^{17}$ erg g$^{-1}\approx 6.7\times 10^{9} L_\odot$ $(M_\odot\mbox{ yr}^{-1})^{-1}$ is the light to mass ratio for a stellar population in statistical equilibrium between star formation and stellar death \citep{Kennicutt2012}.
To presage later discussion,  if the typical IMF varies with environment or evolves with cosmic time or with location because, e.g., of metallicity effects,
then so, too, will the typical $\Psi$.
Moreover, even if the universe realises an approximately constant IMF, because this is itself uncertain \citep[e.g.,][]{Murray2009}, $\Psi$ must itself be considered to be uncertain.
Binarity effects -- not accounted for in this estimate of $\Psi_0$ -- might also be important.
}~
Since the radiation pressure effects we investigate here are expected to have set the final mass configuration within a nascent cluster at 
$t \ll t_{\rm cr}$, i.e., well before its radiative output reaches the statistical equilibrium regime\footnote{On the basis of the modelling by \citet{Skinner2015}
we expect a cluster to have formed and its gas dispersed before $\approx 10 \tau_{\rm ff}$ where $\tau_{\rm ff}$ is the freefall time. 
Conservatively normalising to a cluster of relatively low  volumetric mean density  $\left< \rho_{\star} \right> = 10^5 \msun/(5 \ {\rm pc}^3) \simeq 190 \ \msun {\rm pc}^{-3}$, we have
$\tau_{\rm ff} = [3 \pi/(32 G \rho)]^{1/2} \sim 6 \times 10^5$ yr, so even for this extremal (for our purposes) case we have a formation time $t < t_{\rm cr}$.}, 
we henceforth write 
$L_\star \simeq \Psi M_{\star}$.
The temperature of the cloud surface is
\begin{eqnarray}
T_*  =  \left(\frac{F_\star}{c a} \right)^\frac{1}{4} 
 =  \left(\frac{\Psi  M_{\star}}{ \ 4  \pi R^2 c a} \right)^\frac{1}{4} \, .
\label{eq:Tstar}
\end{eqnarray}
With these equations we can now write the optical depth and Eddington ratio as
\begin{eqnarray}
\tau_* 
& = & \frac{\kappa_{R,*} M_{\rm gas}}{4\pi R^2} 
= \frac{ f_{\rm gas}}{(1 -  f_{\rm gas})} \frac{\kappa_0}{8 T_0^2} \left(\frac{\Psi \Sigma_{\star}^3}{c a}\right)^\frac{1}{2}
\label{eq:tau_star}
 \\
\fE & = &  \frac{\kappa_{R,*} F_\star}{g c} 
 =  \frac{(1 - f_{\rm gas}) \ \kappa_0}{8 \pi G T_0^2} \left(\frac{\Psi^3 \Sigma_{\star}}{c^3 a} \right)^\frac{1}{2}
\label{eq:fsub}
\end{eqnarray}
respectively,
where we have normalised the opacity by defining
\begin{equation}
\kappa_R(T) = \kappa_0 \left(\frac{T}{T_0}\right)^2.
\end{equation}
For Solar metallicity, $\kappa_0 \approx 10^{-1.5}$ cm$^2$ g$^{-1}$ for $T_0 = 10$ K. Note that $M_{\star}$ drops out of both $\tau_*$ and $\fE$; also note that, although the centre-to-edge column density $M/4\pi R^2$ is the quantity that determined $\tau_*$, we have chosen to parameterise our system in terms of the projected column density $\Sigma = M/\pi R^2$, because this is the observable quantity to which we are interested in comparing.

\subsubsection{The stability limit}

\begin{figure}%[hb!]
\centering
\includegraphics[width = \columnwidth]{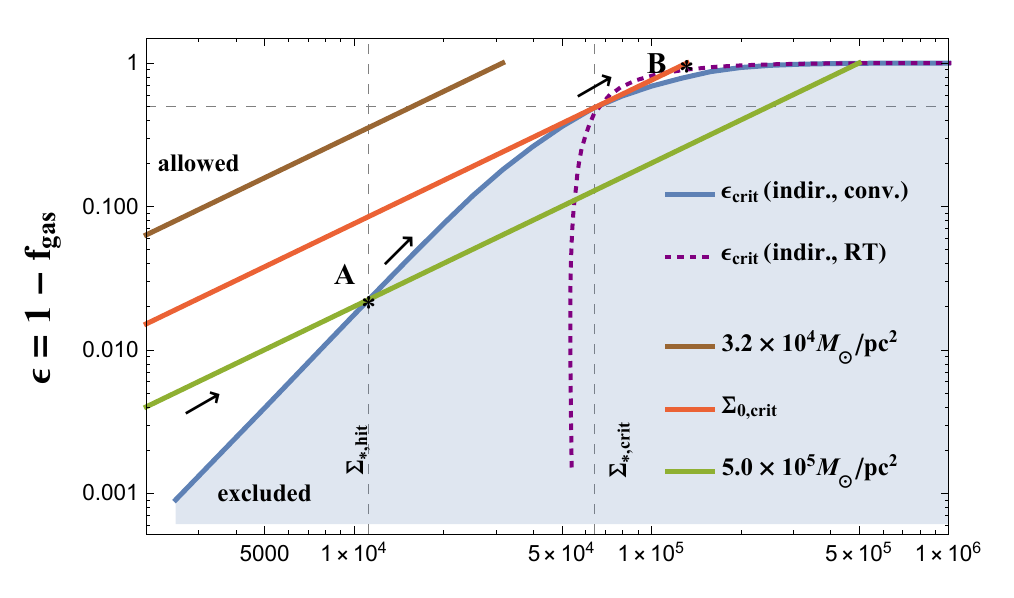}
\includegraphics[width = \columnwidth]{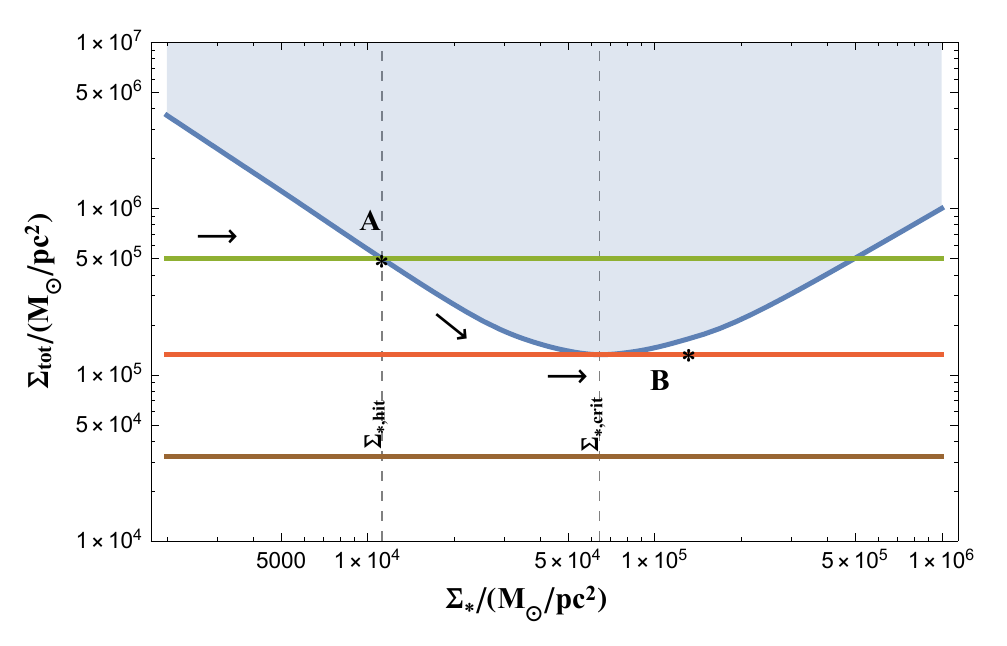}
\caption{
Star formation efficiency $\epsilon$ (top) and total surface density $\Sigma_{\rm tot}$ (bottom) as a function of stellar surface density $\Sigma_{\star}$ for gas of Solar metallicity accounting solely for indirect radiation pressure effects.
The blue curve is defined by the condition $\fE = \fEc$ (\autoref{eq_fEddsub}), with $\fEc$ set equal to the result of \citetalias{Crocker2018}'s case of efficient convection; the shaded blue region, labeled ``excluded'', is the part of parameter space where clouds are unstable to indirect radiation pressure. For comparison, the dashed purple curve shows the separation between stable and unstable assuming no convection. The solid green, red and brown lines show trajectories taken by star-forming clouds with different, fixed total mass surface density ($\Sigma_0$) as indicated; the star formation efficiency and stellar surface densities for these clouds are related by $\epsilon = \Sigma_{\star}/\Sigma_0$, and clouds evolve from lower left to upper right in the top panel as they form stars. Dashed vertical black lines indicate stellar surface densities $\Sigma_{\star,\rm hit}$ (defined in \autoref{sssec:unstable_tracks}) and $\Sigma_{\star,\rm crit}$ (\autoref{eq:Sigma_char}), and the dashed horizontal black line shows $\epsilon=1/2$; the significance of the points `A' and `B' is also described in \autoref{sssec:unstable_tracks}.
\label{fig_plotSubRegionCritfgas}
}
\end{figure}

If we now set 
\begin{equation}
\fE = \fEc(\tau_*, f_{\rm gas}),
\label{eq_fEddsub}
\end{equation}
with $\fE$ evaluated from \autoref{eq:fsub} and $\fEc$ as calculated numerically in \citetalias{Crocker2018} (using \autoref{eq:tau_star} for $\tau_*$), then for any given stellar surface density $\Sigma_{\star}$ we can solve numerically for the gas fraction $f_{\rm gas,crit}$ at which the cloud transitions from stable to unstable against indirect radiation pressure effects. Clouds that have gas fractions $f_{\rm gas} > f_{\rm gas,crit}$ are unstable to indirect radiation pressure, while those with $f_{\rm gas} < f_{\rm gas,crit}$ are stable. Although it is possible to express all our results in terms of gas fraction, it is more intuitive to instead consider a cloud that began its evolution consisting entirely of gas, with surface density $\Sigma_0$, and whose stellar surface density $\Sigma_{\star}$ is a result of subsequent star formation. In this case the gas fraction is simply related to the star formation efficiency $\epsilon$ by 
\begin{equation}
\epsilon = 1 - f_{\rm gas} = \Sigma_{\star}/\Sigma_0 \, . 
\end{equation}
Thus at fixed stellar surface density, clouds that have already processed a large fraction of their gas into stars ($\epsilon \approx 1$) are more stable than those that have processed little gas into stars ($\epsilon \ll 1$), with the two regimes separated at a critical efficiency $\epsilon_{\rm crit} = 1 - f_{\rm gas,crit}$. 
We show these two regimes and the value of $\epsilon_{\rm crit}$ that separates them in \autoref{fig_plotSubRegionCritfgas} (with the uncoloured region marked `allowed' showing the stable zone of parameter space,
and the blue shaded `excluded' region showing the unstable zone); the numerical evaluation used in the plot is for Solar metallicity.

In \autoref{fig_plotSubRegionCritfgas} we also show trajectories in the plane of $(\Sigma_{\star}, \epsilon)$ taken by spherical, star-forming clouds evolving at constant total surface mass densities of $\Sigma_0 = 10^{4.5} \msun$/pc$^2, \Sigma_{\rm 0,crit} \equiv 1.3 \times 10^5 \msun$/pc$^2$ and $5\times 10^{5} \msun$/pc$^2$; we define $\Sigma_{\rm 0,crit}$ in detail below. Since $\Sigma_{\star} = \epsilon \Sigma_0$, in a log-log plot these trajectories are simply lines of slope unity, with a vertical offset determined by $\Sigma_0$. Clouds begin their lives in the lower left corner of the plot, and as they convert gas into stars they move to the upper right. Comparing these cloud trajectories to the stability curve for indirect radiation pressure, it is clear that we can distinguish two types of trajectory. 
Clouds with small initial surface densities, such as the one whose evolutionary path is illustrated by the brown line in \autoref{fig_plotSubRegionCritfgas}, can convert all of their gas into stars without encountering the stability line; indirect radiation pressure effects, therefore, do not limit the gas-to-stars conversion efficiency in such systems.
However direct radiation pressure effects {\it do} limit the eventual stellar surface density as described in \autoref{sssec:unstable_tracks}. 
Clouds with larger values of the initial gas surface density $\Sigma_0$, as illustrated for example by the green line in \autoref{fig_plotSubRegionCritfgas}, will eventually encounter the stability line once they have converted enough of their mass to stars. 
This cloud becomes unstable once in reaches the stellar surface density where the green curve intersects the blue stability line, at the point marked {\bf A} in the plot; we calculate the stellar surface density at which this occurs in the next section. 
These two regimes are separated by a critical initial surface density $\Sigma_0 = \Sigoc \simeq 1.3 \times 10^5 \msun$ pc$^{-2}$ 
(at Solar metallicity),\footnote{This is the value for $\fEc = \fEcc$. 
In the case that $\fEc = \fEcr$, we find $\Sigma_{\rm 0,crit,rt} \simeq 1.2 \times 10^5 \msun$ pc$^{-2}$, nearly identical.
(See \autoref{footnote:CandRT} for the meanings of $\fEcc$ and $\fEcr$.)
} which 
corresponds to $\Sigma_0 = \Sigma_{\rm 0,crit}$, i.e., the value of $\Sigma_0$  such that the cloud trajectory is tangent to the stability curve (shown by 
the red line in \autoref{fig_plotSubRegionCritfgas}  that grazes the blue stability curve at  $\Sigma_\star = \Sigma_{\rm \star,crit}$ indicated by the right vertical dashed line).
We can derive an accurate approximation to $\Sigma_{\rm 0,crit}$  analytically as 
follows.
The first step is to obtain an analytic approximation to the stability curve shown by the blue line in \autoref{fig_plotSubRegionCritfgas}. We do so making use of an analytical approximation derived in \citetalias{Crocker2018}:
\begin{equation}
\fEc \simeq \frac{1 - f_{\rm gas}}{\tau_*}.
\end{equation}
With this approximation, and making use of \autoref{eq:tau_star} and \autoref{eq:fsub}, we can express the stability curve as
\begin{equation}
\Sigma_{\star,\rm crit} \simeq \Sigma_{\rm \star,char} \sqrt{\frac{\epsilon}{1-\epsilon}},
\label{eq:stability_approx}
\end{equation}
where 
\begin{eqnarray}
\label{eq:Sigma_char}
\lefteqn{\Sigma_{\rm \star, char} 
 \equiv \frac{8 \sqrt{\pi G a} \ c {T_0}^2}{\kappa_0 \Psi}} \\
& \simeq & 13.4 \ {\rm g \ cm}^{-2} \left(\frac{\kappa_0}{10^{-1.5} \ {\rm cm}^2\, {\rm g}^{-1}} \right)^{-1}  \left(\frac{\Psi}{\Psi_0}  \right)^{-1}. \nonumber
\end{eqnarray}
Armed with this expression, it is straightforward to find the critical value of $\Sigma_0$ at which the stability curve is tangent to a cloud trajectory $\Sigma_{\star,\rm cloud} = \epsilon \Sigma_0$, since this just amounts to the requirements that $\Sigma_{\rm \star, crit} = \epsilon \Sigma_0$ and
\begin{equation}
\frac{d\Sigma_{\star,\rm crit}}{d\epsilon} = \frac{d\Sigma_{\star,\rm cloud}}{d\epsilon} = \Sigma_0.
\end{equation}
With a bit of algebra one can show that the solution is $\epsilon = 1/2$, and plugging this into the approximation for the stability curve, \autoref{eq:stability_approx}, immediately gives
\begin{equation}
\Sigma_{\rm \star, crit} \simeq \Sigma_{\rm \star, char}
\end{equation}
and
\begin{eqnarray}
\lefteqn{\Sigma_{\rm 0, crit}   =   \Sigma_{\rm \star, crit}/\epsilon \simeq 2 \Sigma_{\rm \star, char} = \frac{16 \sqrt{\pi G a}
\  c {T_0}^2}{\kappa_0 \Psi} }   \nonumber \\ 
&\simeq & 27.6  \ {\rm g \ cm}^{-2} \left(\frac{\kappa_0}{10^{-1.5} \ {\rm cm}^2\,{\rm g}^{-1}} \right)^{-1}  \left(\frac{\Psi}{\Psi_0}  \right)^{-1} \nonumber
\\
&\simeq & 1.3 \times 10^5 \msun \ {\rm pc}^{-2}  \left(\frac{\kappa_0}{10^{-1.5} \ {\rm cm}^2\,{\rm g}^{-1}} \right)^{-1}  
\left(\frac{\Psi}{\Psi_0}  \right)^{-1} \, .
\end{eqnarray}
Assuming -- as is indicated by many observations \citep{Leroy2011,Santini2014,Accurso2017} -- that 
the dust-to-gas ratio scales directly with the gas metallicity at least down to metallicities of $\approx 10\%$ of Solar, and also assuming 
that dust properties are similar to those determined locally, 
these generalise to
\begin{eqnarray}
\Sigma_{\rm 0, crit}(Z) & \simeq & 1.3 \times 10^5 \msun \ {\rm pc}^{-2} \left( \frac{Z}{\zsun} \right)^{-1} \left(\frac{\Psi(Z)}{\Psi_0}  \right)^{-1} \, ,
\end{eqnarray}
where $\Psi = \Psi(Z)$ acknowledges the possibility that there could be metallicity-dependent evolution of the light-to-mass ratio because, e.g., of
systematic IMF changes with $Z$.

\subsubsection{Evolutionary tracks for unstable clouds}
\label{sssec:unstable_tracks}

What happens if $\Sigma_0 > \Sigoc$ as for the green curve in \autoref{fig_plotSubRegionCritfgas}? 
In this case the cloud trajectory will hit the stability curve (as does the example green curve, at the point {\bf A} in \autoref{fig_plotSubRegionCritfgas})  at some stellar surface density $\Sigma_{\rm \star,hit} < \Sigma_{\star, \rm char}$ (indicated by the left vertical line in \autoref{fig_plotSubRegionCritfgas}). 
We can solve numerically for the intersection point $\Sigma_{\rm \star,hit}$, and the corresponding star formation efficiency $\epsilon_{\rm hit}$ at this point in the cloud's evolution, by solving \autoref{eq_fEddsub} simultaneously with $\Sigma_{\star} = \epsilon \Sigma_0$ using our tabulated stability curve $\fEc(\tau_*, f_{\rm gas})$. We can obtain an analytic approximation to this solution using the approximate stability curve given by \autoref{eq:stability_approx}, which yields
\begin{equation}
\Sigma_{\star,\rm hit} = \epsilon_{\rm hit} \Sigma_0 \simeq \Sigma_{\star, \rm char} \sqrt{\frac{\epsilon_{\rm hit}}{1 - \epsilon_{\rm hit}}}.
\end{equation}
The solution is
\begin{eqnarray}
\label{eq:Sigmahit}
\Sigma_{\star,\rm hit} & \simeq & \frac{\Sigoc}{2} \sqrt{\frac{\chi - \sqrt{\chi^2-1}}{\chi+\sqrt{\chi^2-1}}} \\
\epsilon_{\rm hit} & \simeq & \frac{1}{2} \left(1 - \sqrt{1 - \chi^{-2}}\right),
\end{eqnarray}
where $\chi \equiv \Sigma_0 / \Sigoc > 1$. 
The stellar surface density $\Sigma_{\star,\rm hit}$ is the $x-$coordinate of {\bf A} in \autoref{fig_plotSubRegionCritfgas}, and $\epsilon_{\rm hit}$ is the $y-$coordinate of {\bf A}
in the upper panel of \autoref{fig_plotSubRegionCritfgas}.

From this point two evolutionary paths, type A and type B as identified above, are possible.
In the case where a cloud expels its gas and ceases star formation as soon as it becomes unstable (type A), the final efficiency for conversion of gas into stars is simply $\epsilon_f = \epsilon_{\rm hit}$, and the corresponding stellar surface density is $\Sigma_f = \epsilon_f \Sigma_0$; the end point of such an evolutionary history is indicated by the asterisk labelled `A' in \autoref{fig_plotSubRegionCritfgas}. The second evolutionary path (type B) is one wherein the cloud expands in order to maintain marginal stability, and is illustrated by the arrows in \autoref{fig_plotSubRegionCritfgas}. In this case a cloud begins evolving at constant $\Sigma_{\rm tot}$, and thus along a line of of slope unity described by $\epsilon = \Sigma_{\star}/\Sigma_0$. However, once the cloud hits the stability line at $\Sigma_{\star,\rm hit}$, it cannot continue to evolve at constant $\Sigma_{\rm tot}$. 
Instead, it travels \textit{along} the marginal stability curve, which corresponds to expanding in radius while the star formation efficiency increases. It remains on this curve, at the edge of stability, until reaching the tangent point at $\Sigma_{\star} \simeq \Sigma_{\star, \rm char}, \epsilon \simeq 1/2$, at which point it can continue its evolution at a new constant surface density $\Sigma_{\rm tot} = \Sigoc$. 
If indirect radiation pressure were the only effect limiting star formation, the cloud would then proceed to form stars at this surface density until it converted all its mass to stars and reached a final star formation efficiency $\epsilon_f = 1$; the endpoint of such a trajectory is indicated by the `B' in \autoref{fig_plotSubRegionCritfgas}. In this case the final stellar surface density is simply $\Sigma_f = \Sigoc$. We will see in the next section that the effects of direct radiation pressure alter this conclusion, though only slightly.\footnote{Note here that we are defining final stellar surface density simply as the total mass in stars formed divided by the final cross-sectional area. In reality, for a type B evolutionary path stars that form before the cloud becomes unstable to indirect radiation pressure could form in a more compact configuration than stars that form later, after the cloud has begun expanding. In principle we could use a more sophisticated definition of surface density that weights by mass formed in different phases of the evolution. However, such an estimate could differ from our simple one by at most a factor of $\approx 2$, because half of the total mass in stars (or slightly less once direct radiation pressure is included) forms after the cloud reaches a total surface density $\Sigoc$. Given the simplicity or our model overall, the extra algebraic complexity required to capture this factor of 2 effect does not seem warranted.}

\subsection{Evolutionary tracks with direct radiation pressure}
\label{ssec:tracks_full}

\begin{figure}%[hb!]
\centering
\includegraphics[width = \columnwidth]{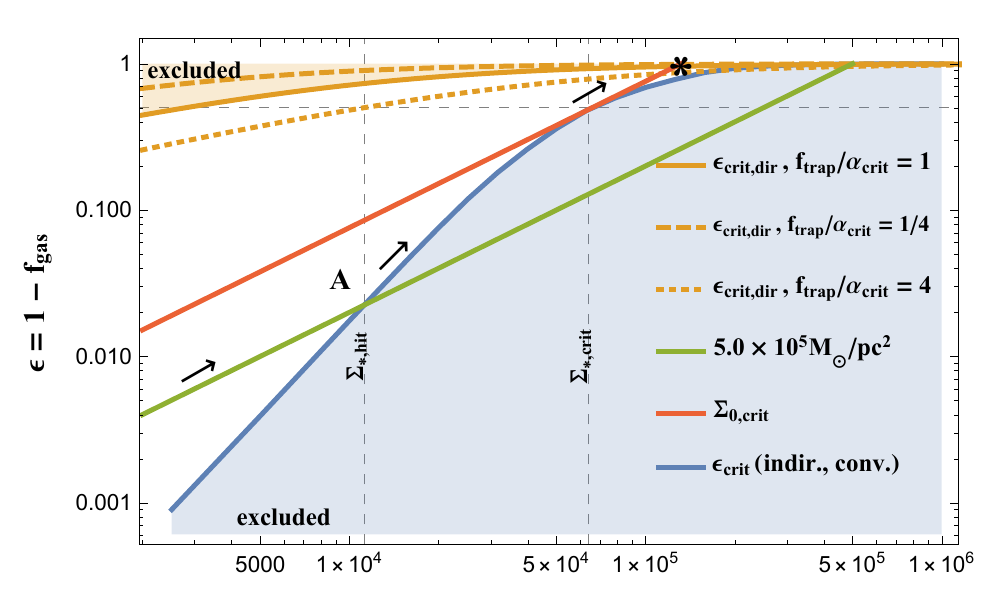}
\includegraphics[width = \columnwidth]{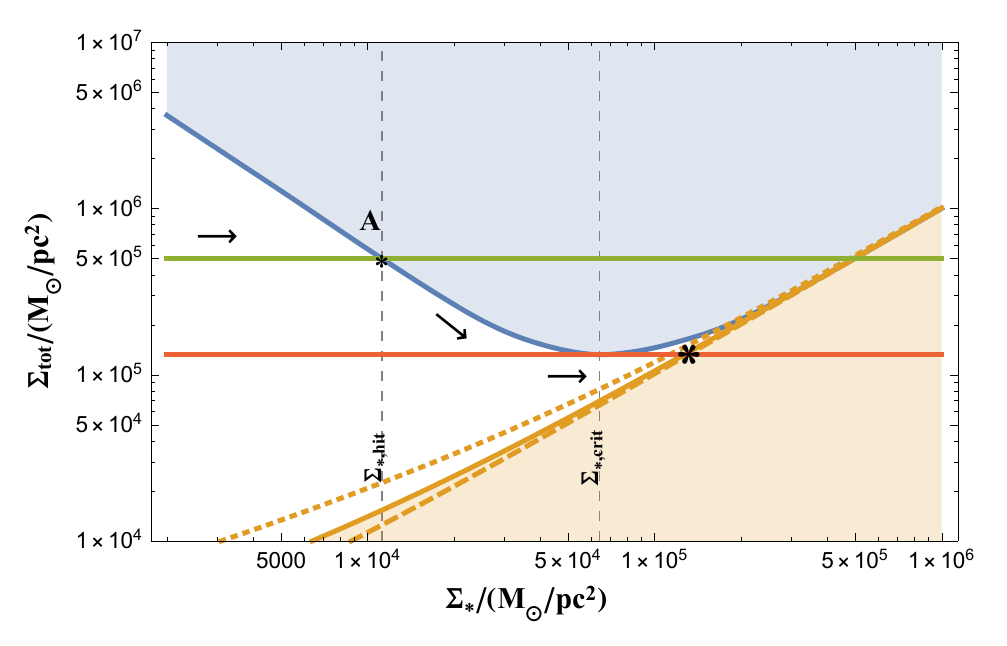}
\caption{
Star formation efficiency $\epsilon$ (top) and total surface density $\Sigma_{\rm tot}$ (bottom) as a function of stellar surface density $\Sigma_{\star}$, incorporating both direct and indirect radiation pressure effects. 
The plots adopt $\Psi = \Psi_0$.
The blue shaded region shows the locus within which clouds are unstable against indirect radiation pressure (for Solar metallicity), while the yellow shaded region shows the locus of instability against direct radiation pressure (for $\eta = 2/3$ and the indicated values of $f_{\rm trap}/\alpha_{\rm crit} = 1$). The green line shows an example trajectory at fixed $\Sigma_0 = 5\times 10^5 \msun$/pc$^2 > \Sigma_{\rm 0,crit}$, while the red line shows an example trajectory with $\Sigma_0 = \Sigoc$. Arrows show the path taken by a cloud that expands while forming stars so as to remain marginally-stable against indirect radiation pressure.
} 
\label{fig_plotEpsilonIndirPlusDir}
\end{figure}

As presaged above, we next incorporate {\it direct} radiation pressure effects following \citetalias{Fall2010}.
For direct radiation pressure, the critical luminosity at which gas is expelled corresponds to that which delivers enough momentum within one dynamical time to drive
the atmosphere to roughly the escape speed. \citetalias{Fall2010} show that the required luminosity is
\begin{equation}
L_{\rm crit,dir} = \alpha_{\rm crit} \frac{G c (1 - \epsilon) M_{\rm tot}^2}{5 \eta f_{\rm trap} R^2},
\label{eq:L_crit}
\end{equation}
where $f_{\rm trap} \sim 2-5$ accounts for assistance from main-sequence winds and incomplete leakage of starlight and wind energy \citep{Krumholz2009}, $\alpha_{\rm crit}$ is a parameter of order unity that accounts for magnetic support and turbulent support, and $\eta = 2/(4 - k)$\footnote{Note that $k > -4$ is required for the similarity solution presented by \citet{Krumholz2009}; a different similarity solution for $k \leq -4$ can  be obtained, but such a steep density profile does not seem physically relevant and we do not consider it further.} where the internal density profile of the cloud follows $\rho \propto r^{-k}$; following \citetalias{Fall2010} we adopt a fiducial $\eta = 2/3$ and consider $f_{\rm trap}/\alpha_{\rm crit}$ in the range $1/4 - 4$. We can rewrite this critical luminosity in terms of a flux at the surface cloud surface by dividing both sides by $4\pi R^2$:
\begin{eqnarray}
F_{\rm crit,dir} & = & \alpha_{\rm crit} \frac{\pi}{20} \frac{G c}{\eta f_{\rm trap}} f_{\rm gas} \Sigma_{\rm tot}^2 \nonumber \\
& = & \alpha_{\rm crit} \frac{\pi}{20} \frac{G c}{\eta f_{\rm trap}} \frac{f_{\rm gas}  \Sigma_{*}^2}{(1 -  f_{\rm gas})^2} \, .
\end{eqnarray}
Now equating the stellar radiation flux $F = \Psi M_{\star}/(4 \pi R^2) = \Psi \Sigma_{\star}/4$ to the critical flux we obtain a critical stellar surface density required for direct radiation pressure force to eject the gas and halt star formation:
\begin{eqnarray}
\Sigma_{\rm *,crit,dir} & = & \frac{5}{\pi} \frac{\eta f_{\rm trap} \Psi}{\alpha_{\rm crit} G c} \frac{(1 - f_{\rm gas})^2}{f_{\rm gas}} \nonumber \\
& = & \frac{5}{\pi} \frac{\eta f_{\rm trap} \Psi}{\alpha_{\rm crit} G c} \frac{\epsilon^2}{1-\epsilon} \, .
\label{eq:Sigmacrit}
\end{eqnarray}
Defining a characteristic scale for stellar surface density due to direct radiation pressure effects  
\begin{eqnarray}
\Sigma_{\rm \star,char,dir} & \equiv & \frac{5}{\pi} \frac{\eta f_{\rm trap} \Psi}{\alpha_{\rm crit} G c} \nonumber \\
& \simeq & 5900 \ \msun\,{\rm pc}^{-2} \left(\frac{\eta}{2/3}\right) 
\left(\frac{f_{\rm trap}}{\alpha_{\rm crit}}\right) \left( \frac{\Psi}{\Psi_0} \right) \, ,
\end{eqnarray}
we can solve \autoref{eq:Sigmacrit} to find (taking the positive root) 
\begin{equation}
\epsilon_{\rm crit,dir}(\Sigma_{\star}) = \frac{\Sigma_{\star}}{2 \Sigma_{\rm \star,char,dir}} \left[ \left(1 + \frac{4 \Sigma_{\rm \star,char,dir}}{\Sigma_{\star}} \right)^{\frac{1}{2}} -1 \right] 
 \, ;
\label{eq:eps_dir}
\end{equation}
\citepalias[this equation is equivalent to][Eq.~6]{Fall2010}.
Some example curves for $\epsilon_{\rm crit,dir}(\Sigma_{\star})$ are shown in yellow in \autoref{fig_plotEpsilonIndirPlusDir}, with the shaded yellow region indicating the part of parameter space where clouds are unstable to disruption by direct radiation pressure.

Note that, unlike for indirect radiation pressure, once the $(\Sigma_{\star},\epsilon)$ configuration of the system crosses the $\epsilon_{\rm crit,dir}(\Sigma_{\star})$
critical line, direct radiation pressure invariably blows out the remaining gas in the system, cutting off further star formation. The cloud cannot stabilise by expanding as it can for indirect radiation pressure, because the radial dependence of the radiation forces is fundamentally different. For indirect radiation pressure, expansion causes the radiation force per unit mass to drop faster than $R^{-2}$, because the flux falls as $R^{-2}$ and the opacity falls as the cloud expands as well. For direct radiation forces, on the other hand, the extremely high opacity of dusty interstellar gas to radiation with the colour temperature of a star ($\kappa \sim 10^3$ cm$^2$ g$^{-1}$) ensures that all photons will be absorbed for any reasonable column of gas. Consequently, the force delivered by the direct radiation field is simply the total momentum flux in the radiation field, which is constant and independent of radius \citep[cf.][]{Thompson2015}. Thus expansion makes gravity stronger compared to the indirect radiation force, but weaker compared to the direct force, and for direct radiation forces expansion is therefore destabilising rather than stabilising.

Examining \autoref{fig_plotEpsilonIndirPlusDir}, it is clear that clouds that begin their evolution with surface densities $\Sigma_0 < \Sigoc$ (i.e., with trajectories above the red line in the upper panel and below the red line in the lower panel) will be unbound by direct radiation pressure without ever being unstable against indirect radiation pressure, while clouds with $\Sigma_0 > \Sigoc$ (with an example shown by the green curve in upper and lower panels of \autoref{fig_plotEpsilonIndirPlusDir}) will either be disrupted by indirect radiation pressure (under our `type A' assumption) or will be forced to expand by indirect radiation pressure to surface density $\Sigoc$ (under our `type B' assumption), after which the cloud will continue to form stars at constant surface density until it is disrupted by direct radiation pressure.\footnote{For a sufficiently large value of $f_{\rm trap}/\alpha_{\rm crit}$, or a sufficiently high metallicity, it is possible that the tangent point on the indirect radiation pressure curve could move within the zone of instability for direct radiation pressure, in which case clouds undergoing expansion under the effects of indirect radiation pressure would become vulnerable to disruption by direct radiation pressure before reaching a surface density $\Sigoc$. Since this case would require either very large values of $f_{\rm trap}$ or highly super-Solar metallicities, it seems unlikely to be relevant in nature, and thus we do not discuss it further.} We show examples of both types of cloud trajectories in \autoref{fig_plotEpsilonIndirPlusDir}.

For type B evolutionary trajectories, where the cloud expands if it becomes unstable to indirect radiation pressure, the final star formation efficiency and surface density are always dictated by the total surface density of the cloud when it encounters the line of direct radiation pressure instability; this is $\Sigma_{\rm tot} = \min(\Sigma_0, \Sigoc)$, with the first case corresponding to clouds that begin their evolution at low surface density, $\Sigma_0 < \Sigoc$, and the latter to clouds that begin their lives at high surface density, $\Sigma_0 > \Sigoc$. In either case we can calculate the final star formation efficiency and stellar surface density analytically, simply by making the substitution $\Sigma_{\star} = \epsilon_f \Sigma_{\rm tot}$ in \autoref{eq:eps_dir} and solving. The result is
\begin{eqnarray}
\epsilon_f & = & \frac{1}{1 + \Sigma_{*,\rm char,dir}/\min\left(\Sigma_0, \Sigoc\right)} \\
\Sigma_f & = & \frac{\min\left(\Sigma_0,\Sigoc\right)}{1 + \Sigma_{*,\rm char,dir}/\min\left(\Sigma_0,\Sigoc\right)}.
\end{eqnarray}
The maxima of these two functions occurs for $\Sigma_0 \geq \Sigoc$, and for our fiducial parameter choices and Solar metallicity, these maxima are $\epsilon_f \simeq 0.88$ and $\Sigma_f \simeq 1.3\times 10^5$ $M_\odot$ pc$^{-2}$. For type B trajectories, since the surface density of the final stellar state is different to that of the initial gaseous state, it is useful to define an additional measure of efficiency,
\begin{equation}
\epsilon_{f,\Sigma} = \frac{\Sigma_{\star}}{\Sigma_0} = \frac{\min\left(1, \Sigoc/\Sigma_0\right)}{1 + \Sigma_{*,\rm char,dir}/\min\left(\Sigma_0,\Sigoc\right)}.
\end{equation}
Intuitively, the difference between $\epsilon_f$ and $\epsilon_{f,\Sigma}$ is that the former measures the fraction of the initial gas mass transformed into stars, while the latter measures the ratio of the final stellar surface density to the initial gas surface density; thus the latter quantity accounts for the effects of expansion. The maximum possible value of $\epsilon_{f,\Sigma}$ is the same as for $\epsilon_f$.

For type A trajectories, where we assume clouds disrupt instantaneously if they become unstable to direct radiation pressure, the results for $\epsilon_f$ and $\Sigma_f$ are identical to the type B case for $\Sigma_0 < \Sigoc$, while for $\Sigma_0 > \Sigoc$ the final stellar surface density and star formation efficiency are given by the solution described in \autoref{sssec:unstable_tracks}.

\begin{figure}%[hb!]
\centering
\includegraphics[width = \columnwidth]{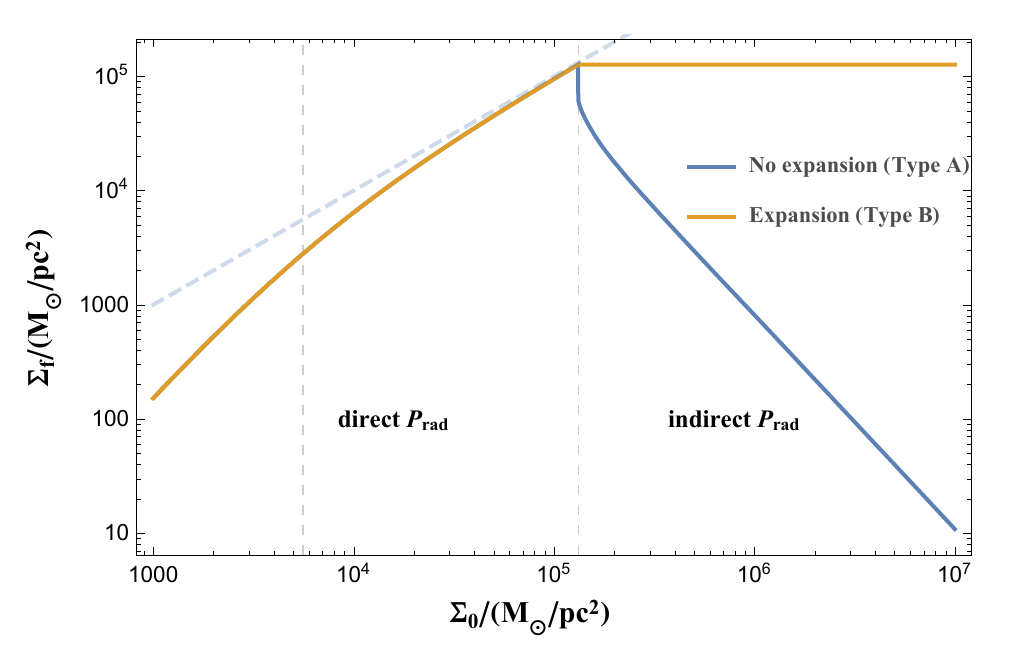}
\caption{
Final stellar surface density $\Sigma_f$ as a function of initial mass surface density of the star forming cloud $\Sigma_0$ for  $\Psi = \Psi_0$.
At the low mean surface density end, direct radiation pressure effects (in this case, for fiducial parameters of $\eta = 2/3$ and $f_{\rm trap}/\alpha_{\rm crit}  = 1$) limit the final stellar surface density; the left dashed vertical line indicates $\Sigma_{*,\rm char,dir}$ the surface density for which the star formation efficiency permitted by direct radiation pressure is 50\%.
At the high surface mass density end, indirect radiation pressure effects (in this plot calculated for Solar metallicity) constrain the final stellar surface density. 
The {\it blue} curve assumes that the cloud is disrupted when its surface density trajectory intersects the critical curve for 
indirect radiation pressure (type A trajectory); the {\it yellow} curve assumes that the cloud can expand dynamically to maintain marginal stability.
The dashed grey curve shows $\Sigma_f = \Sigma_0$.
In either case, the model predicts a maximal stellar surface density at $\Sigma_{\star} \sim \Sigoc$, indicated by the right dashed vertical line, though note the expansion case predicts a pile-up at this value whereas the no-expansion case predicts a peak with a downturn on either side of it.
} 
\label{fig_plotFinalVsInitSurfaceDensity}
\end{figure}

\begin{figure}%[hb!]
\centering
\includegraphics[width = \columnwidth]{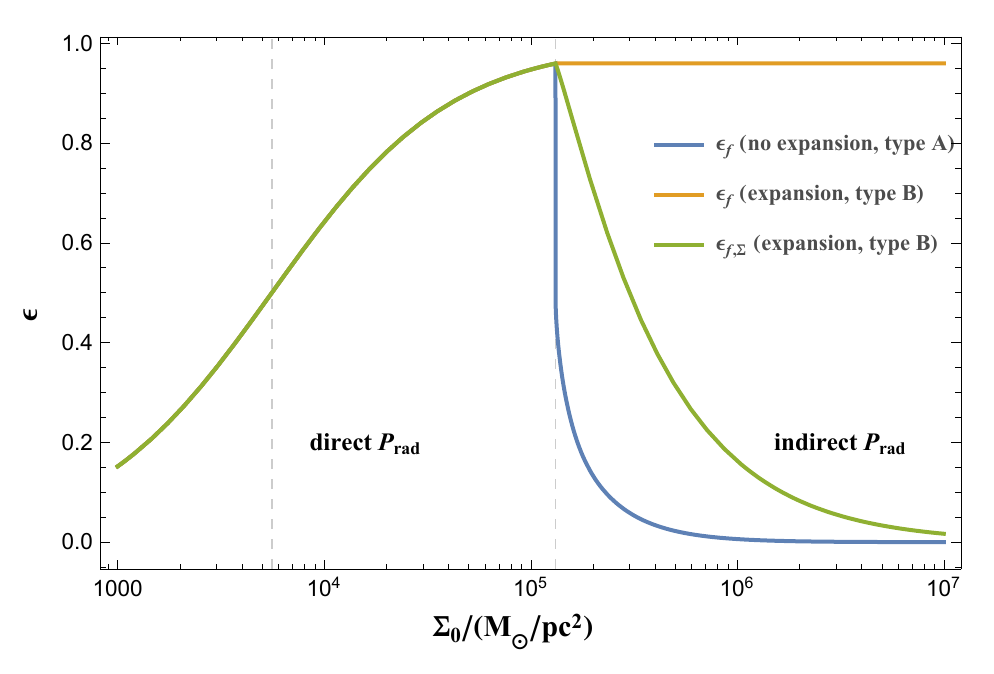}
\caption{
Final star-formation efficiency, defined by mass ($\epsilon_f$) or by surface density ($\epsilon_{f,\Sigma}$) as indicated in the legend, as a function of initial cloud mass surface density, $\Sigma_0$; the curves are calculated using the same parameter values as \autoref{fig_plotFinalVsInitSurfaceDensity}, and the dashed vertical lines are the same in both cases as well.
For the no expansion (type A) case, and for the expansion case with starting surface density $\Sigma_0 < \Sigoc$, both efficiency measures are the same, because the cloud radius does not change during star formation.
} 
\label{fig_plotFinalEfficiencyDisplay}
\end{figure}

We show the relationship between $\Sigma_0$ and $\Sigma_f$ for both cases in \autoref{fig_plotFinalVsInitSurfaceDensity}, and we show the corresponding relationship between $\Sigma_0$ and $\epsilon_f$ (or $\epsilon_{f,\Sigma}$) in \autoref{fig_plotFinalEfficiencyDisplay}. In these plots we use the numerical value for $\Sigoc$ rather than the analytic approximation, although the two differ only very slightly.

\section{Discussion}
\label{sec:discussion}

\subsection{Limits on surface density: observational comparison}

The simple considerations set out above suggest that indirect and direct radiation pressure, acting in concert, should put an upper limit of 
$\sim 1.3 \times 10^5 \msun$pc$^{-2} \ (\zsun/Z) \ (\Psi_0/\Psi)$ for the maximum stellar surface density of star clusters.
Interestingly, this finding is broadly consistent with observations.
Indeed, as presaged above, \citet{Hopkins2010} pointed out the apparent existence of an  empirical upper bound, at $\Sigma_{\rm max} \sim 10^5  \msun$pc$^{-2}$, to the  
central stellar surface mass density of compact stellar systems.
 \citet{Hopkins2010}  remark that this rough limit holds over a $\sim$ 7 order magnitude range of total mass, a $\sim$ 5 order of magnitude range
in physical size, and over $\sim 2$ orders of magnitude in metallicity.
Given this, they tentatively conclude  ``that feedback from massive stars likely accounts for the observed  $\Sigma_{\rm max}$, plausibly because star formation reaches an Eddington-like flux that regulates the growth of these diverse systems".
However, a problem for this explanation -- raised by \citet{Hopkins2010} and again recently by \citet{Grudic2018a,Grudic2018b} -- is the expectation that 
$\Sigma_{\rm max}$ should scale inversely as the metallicity (because of the expectation that the dust-to-gas mass ratio trace metallicity) whereas it did not seem to these authors that there was any such metallicity dependency evident in their data.

\begin{figure}%[hb!]
\centering
\includegraphics[width = 0.5 \textwidth]{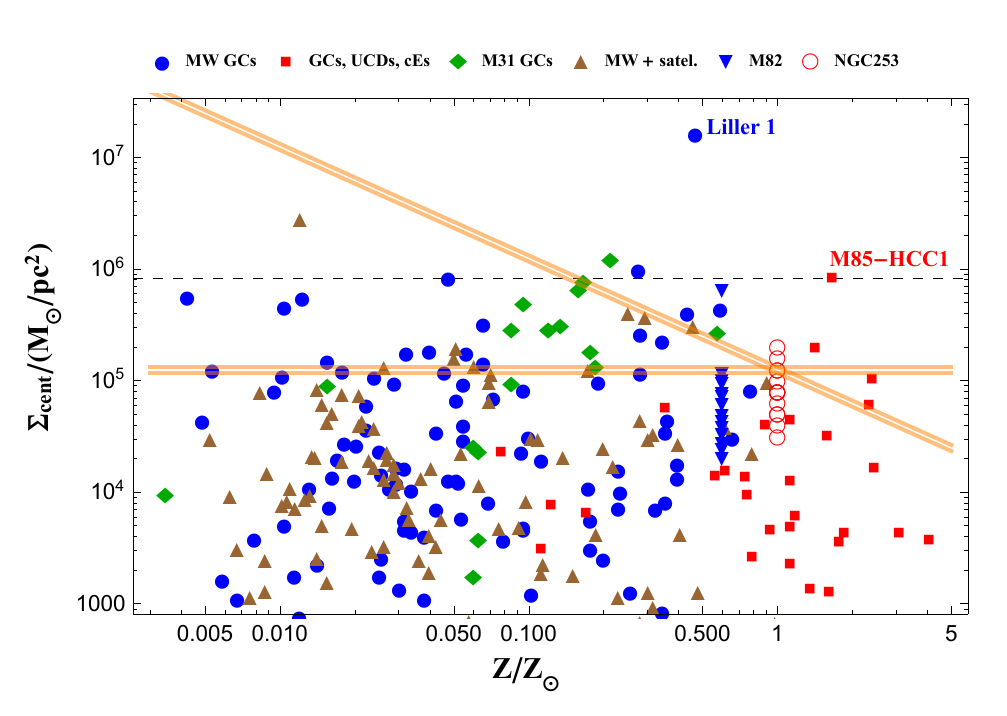}
\caption{
Central surface mass density (measured, inferred, or lower limit as described below) versus metallicity for different compact stellar systems (data truncated below $10^3 \msun$ pc$^{-2}$). 
The systems included in the compilation are as follows:
{\bf MW GCs:} Milky Way globular clusters ($N = 119$) with central mass surface density from 
\citet{Baumgardt2018} and with metallicities from \citet{Harris1996,Harris2010};
{\bf GCs, UCDs, cEs:} surface mass density within the effective half-light radius for massive globular clusters, ultracompact dwarfs, and
compact elliptical galaxies ($N = 29$) from the  Archive of Intermediate Mass Stellar Systems with metallicities compiled by \citet{Janz2016};
{\bf M31 GCs:} Central surface mass densities for globular clusters from M31 ($N = 18$) from the compilation by \citet{Barmby2007};
{\bf MW + satel.:} Central surface mass densities for massive clusters from Milky Way and satellites 
(the Large Magellanic Cloud, Small Magellanic Cloud and the Fornax dwarf spheroidal, $N = 153$) from the compilation by \citet{McLaughlin2005};
{\bf M82:} surface density inside half mass radius for super star clusters from M82 from \citet{McCrady2007} ($N = 15$) adopting gas phase metallicity of 0.6 $\zsun$ \citep{Origlia2004,Nagao2011};
{\bf NGC253:} total (gas+stars) surface density inside 2D FWHM radius nascent super star clusters from the central starburst in NGC 253 from \citet{Leroy2018} ($N = 14$) adopting gas phase metallicity of 1.0 $\zsun$ \citep{Webster1983}.
The sloped orange line is for  $\Sigma_{\rm 0,crit}(Z) = \Sigma_{\rm 0,crit}(\zsun/Z)$ with $\Psi = \Psi_0$ and 
a width encompassing the efficient convection and the pure radiative transfer limits
(the horizontal orange line is for no metallicity evolution).
The dashed horizontal line corresponds (cf.~\autoref{eq:SigmaCritIMBH}) \ roughly to the critical stellar volumetric number density of $n_* \sim 10^6$ pc$^{-3}$ where stellar mergers
in cluster cores occur on a timescale less than the main sequence lifetime of massive stars, allowing 
the merger formation of giant stars, progenitors of intermediate mass black holes
(assuming a large fraction of massive stars are in primordial hard binaries; see the text).
The two systems that most surpass the expected upper limit are labelled; they are M85-HCC1, a hypercompact star cluster claimed as likely the remnant nucleus of a galaxy recently accreted on to its current host galaxy \citep{Sandoval2015}, and Liller 1, a Galactic globular cluster that is in core collapse \citep{Baumgardt2018}.
} 
\label{fig_plotCentralSurfaceDensityVsMetallicity.pdf}
\end{figure}

Our compilation of the central surface mass density of a number of systems vs.~metallicity in shown in \autoref{fig_plotCentralSurfaceDensityVsMetallicity.pdf}.
The data are still sparsely sampled and subject to selection effects that we have not rigorously characterised.
Nevertheless, 
they do not seem to us to  be inconsistent with such a metallicity-dependency for $\Sigma_{\rm max}$.

A further point to keep in mind here is that the sloped limit line in \autoref{fig_plotCentralSurfaceDensityVsMetallicity.pdf} assumes a metallicity independent light to mass ratio $\Psi(Z) \to \Psi_0$.
Were, for instance, the mean IMF to become systematically more ``top heavy" for lower metallicity, 
this would have the overall effect of attenuating and plausibly even totally cancelling out the metallicity evolution of the critical central surface mass density \citep[cf.~recent observations of the 30 Doradus local starburst:][where the expected $\sim 3-4$ increase in $L/M$ concomitant with the claimed top-heavy IMF would roughly cancel out the effect of the LMC's $\sim 0.4 \zsun$]{Schneider2018}.
However we caution the reader that  claims for strong metallicity effects on the IMF have frequently been contested or countered with seemingly contradictory examples or analysis \citep[e.g.,][]{Bastian2010, Offner2014}.

Aside from obtaining more data, a fuller empirical analysis of the issue of the putative maximal central surface mass density would have to deal with a number of issues that revolve around questions of systematics connected to the dynamical evolution of stellar systems.
One possible confound is that metal-rich GCs tend to have formed nearer the centres of their host galaxies than relatively metal poor ones, i.e., in a relatively stronger tidal fields. 
This will render the former more compact at the present time, independent of their initial configuration.

In light of the complexity of applying our model to old stellar systems that may have undergone significant dynamical evolution, the data covering in-formation, embedded super star clusters in the NGC 253 nuclear star burst obtained by \citet{Leroy2018} provide a particularly interesting test of our model.
From ALMA 350 MHz dust continuum observations, these authors find 14 candidate super star clusters ($M_{\star} \gsim 10^5 \msun$) within the $\sim 3 \times 10^8 \msun$ of molecular gas inside NGC 253's nucleus.
Each cluster has a large gas fraction and is, consequently, highly extincted behind very high gas and dust columns (yielding optical depths $\tau \sim 5 -10$ at 100 $\mu$m).
Nevertheless, \citet{Leroy2018} find no high velocity line wings in their data implying that each cluster's gas is gravitationally bound.
This finding is broadly consistent with our scenario which suggests that indirect radiation pressure effects in the nascent NGC 253 star clusters, 
while likely to cause some of the clusters to expand, 
will not expel the clusters' gas (cf.~\autoref{fig_plotCentralSurfaceDensityVsMetallicity.pdf}).
Overall, indeed, we would expect a large fraction of each cluster's original gas allocation to be turned into stars, consistent with the data and analysis presented by \citet{Leroy2018}.

\subsection{Stellar Mergers and Intermediate Mass Black Holes}

A  possible consequence of our picture is that clusters formed at higher metallicities (approaching solar) may be prevented by indirect radiation pressure from reaching the volumetric stellar number densities required such that stellar mergers may occur sufficiently rapidly in their cores to form the giant stellar precursors of intermediate mass black holes (IMBHs).
\citet{Bonnell2005} find that stellar mergers of massive stars can occur on a timescale less than their $\sim 10^6$ yr main sequence lifetimes -- thus allowing for the dynamical formation of giant stars via mergers  -- for core cluster densities $n_* \gsim 10^6$ pc$^{-3}$ assuming that a large fraction of massive stars are born into hard binary systems.
This corresponds to a rough critical surface density
\begin{equation}
\Sigma_{\rm crit,IMBH} \sim 8 \times 10^5 \msun \ {\rm pc}^{-2} \left(\frac{M_c}{10^6 \msun} \right)^{\frac{1}{3}} \left(\frac{\rho_{\rm crit,IMBH}}{10^6 \msun/{\rm pc}^3} \right)^{\frac{2}{3}} \, .
\label{eq:SigmaCritIMBH}
\end{equation}
This critical surface density is shown, for the fiducial parameters, as the horizontal dashed line in
\autoref{fig_plotCentralSurfaceDensityVsMetallicity.pdf}; compact stellar systems at $Z \gsim 0.2 \zsun$ may have trouble reaching this critical surface density (though it is to be acknowledged that there are a handful of star systems above this rough threshold at $Z \gsim 0.2 \zsun$ in our compilation).
This effect is independent of and in addition to stellar mass loss due to Wolf-Rayet phase winds which also increases as a function of metallicity and may also  preclude the formation of IMBHs \citep{Yungelson2008,Glebbeek2009}.

\section{Summary}
\label{sec:summary}

In this paper we investigate the combined effects of indirect, dust-reprocessed and direct, stellar radiation pressure on the formation of the densest star clusters. 
Adopting our calculation from \citetalias{Crocker2018} of the largest  column of dusty gas for which hydrostatic equilibrium  is possible when the column is subject to the opposing forces of indirect radiation pressure and gravity and also drawing results on direct radiation pressure  from \citetalias{Fall2010}, 
we construct here an evolutionary model for a dense protocluster gas cloud forming stars, and show that clouds with initial surface density $\gsim 10^5  \msun$ pc$^{-2}$ with Milky Way-like dust and dust-to-gas ratios will become super-Eddington with respect to indirect radiation pressure at some point during their star formation process. 
This effect is unlikely to push gas out of clusters in winds, abruptly cutting-off star formation.
Rather, systems with sufficiently high initial surface mass densities probably suffer a rather gentle expansion under indirect radiation pressure effects.
Only late in their evolution, once they have typically turned well more than 50\% of their original gas allocation into stars, will {\it direct} radiation pressure effects turn on in such clusters, 
pushing the remaining gas out of the cluster at greater than the escape speed.
This process is complete well before core-collapse supernovae start going off in such clusters.

The  combined effect of direct and indirect radiation pressure is to set an upper limit of $\approx 10^5 \ (Z_\odot/Z)$ $M_\odot$ pc$^{-2}$ on the surface densities of star clusters, and to produce a star formation efficiency -- defined as the ratio of the initial gas surface density to the final stellar surface density -- that has a maximum of $\approx 90\%$ for gas clouds with surface densities near this upper limit, and falls off sharply at either lower or higher surface densities. This limit is likely to preclude the formation of intermediate mass black holes via stellar collisions in any star cluster with a metallicity above $\approx 20\%$ of Solar. The scenario suggested by our model is qualitatively and quantitatively consistent with the empirical determination \citep{Hopkins2010} that there seems to be an upper limit to the central surface mass density of compact stellar systems at $\sim 10^5 \msun$ pc$^{-2}$. 
We compare our model to an updated compilation of measured stellar surface densities in a wide variety of compact stellar systems, and find that it is qualitatively consistent with the observations, including the possibility that star clusters formed with lower metallicities and thus lower dust content might have systematically higher maximum surface densities. Further tests of this model, and in particular the prediction that the results should depend on metallicity, will require additional measurements of surface densities in low-metallicity systems, preferably young ones so as to minimise the confounding effects of dynamical evolution.

\section*{Acknowledgements}

The authors 
thank Mike Grudi\'c and Phil Hopkins for helpfully providing a compilation of stellar cluster numerical data.
This research has made use of the VizieR catalogue access tool, CDS,
 Strasbourg, France. The original description of the VizieR service was
 published in A\&AS 143, 23.
MRK acknowledges support from the Australian Research Council's Discovery Projects grant DP160100695 and the ARC Centre of Excellence for All Sky Astrophysics in 3 Dimensions, project CE170100013. 
TAT is supported in part by NSF \#1516967 and by NASA ATP 80NSSC18K0526.
DM acknowledges support from the Australian Research Council Future Fellowship FT160100206.

%%%%%%%%%%%%%%%%%%%%%%%%%%%%%%%%%%%%%%%%%%%%%%%%%%



%\bibliographystyle{mnras}


%%%%%%%%%%%%%%%%%%%%%%%%%%%%%%%%%%%%%%%%%%%%%%%%%%

%%%%%%%%%%%%%%%%% APPENDICES %%%%%%%%%%%%%%%%%%%%%

%\begin{appendix}
%\appendix
\begin{appendices}

%\section{Analytic derivation of $\Sigma_{\rm 0,crit}$}
%\label{sec:SigmaCritDrvtn}

\section{Validity of the $\kappa_R \propto T^2$ approximation}
\label{sec:T0appendix}

\begin{figure}%[hb!]
\centering
\includegraphics[width = 0.5 \textwidth]{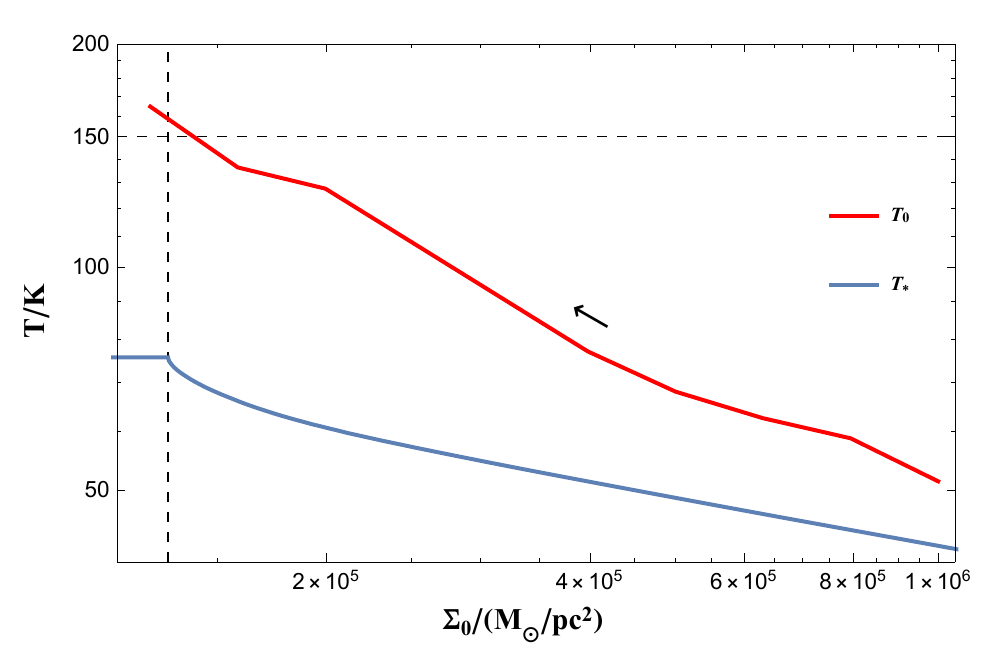}
\caption{Reference ($T_*$) and central temperature as a function of total surface density for marginally-stable gas clouds. The central temperature is as calculated in \citetalias{Crocker2018} for planar geometry.
The dashed horizontal line at $T = 150$ K indicates where $\kappa_R(T)$ starts to deviate from a simple $\propto T^2$ behaviour \citep{semenov03a}.
The vertical dashed line indicates $\Sigma_{\rm 0,crit}$, the lowest total surface mass density where indirect radiation pressure effects manifest.
As indicated by the arrow, systems with $\Sigma_{\rm 0} > \Sigma_{\rm 0,crit}$ evolve from right to left as they process gas into stars while simultaneously expanding (with their core gas heating up). 
The plot shows that such systems remain in the regime $T < 150$ K over almost the entirety of their evolution.
} 
\label{fig_plotTempAtSigmastarHit}
\end{figure}

Our calculation of the critical Eddington ratio $f_{\rm E,crit}$ demarcating stable, sub-Eddington from unstable, super-Eddington behaviour assumes that the temperature $T \lesssim 150-200$ K, so that we can approximate the temperature-dependence of the opacity as $\kappa_R \propto T^2$. Here we verify that this is in fact the case.
In \autoref{fig_plotTempAtSigmastarHit} we plot the reference temperature $T_*$ (\autoref{eq:Tstar}) and the central (i.e., midplane) temperature as calculated in \citetalias{Crocker2018} as a function of surface density for marginally stable systems, i.e., for gas clouds with total surface density $\Sigma_0$ and gas fraction
$f_{\rm gas, crit} = 1 - \epsilon_{\rm crit}(\Sigma_0)$.
Note that the $\kappa_R \propto T^2$ behaviour of the opacity breaks down in the temperature range 150-200 K \citep{semenov03a}, indicated in \autoref{fig_plotTempAtSigmastarHit} by the horizontal dashed line.
Systems with $\Sigma_{\rm 0} > \Sigma_{\rm 0,crit}$ evolve from right to left as they process gas into stars while simultaneously expanding.
This plot indicates that, to the extent that it is accurate to treat the geometry in the planar limit,
a $\propto T^2$ scaling of $\kappa(T)$ is a very good approximation for most of the evolution of the forming clusters, only (slightly) breaking down as systems approach $\Sigma_{\rm 0,crit}$.

On very small scales around individual stars or stellar systems, of course, the planar approximation must break down, and the radiation temperatures will be higher. However, the scales on which this is the case are \textit{very} small. For example, \citet{Wolfire1986} show that, even around a 300 $M_\odot$ star with a luminosity of $7\times 10^6$ $L_\odot$, the dust temperature exceeds 200 K only within $\sim 10^3$ AU of the source. Even for the very dense star clusters we are considering, this is smaller than the typical separation for such luminous sources. For example, there are $\sim 50$ early O and WR stars that approach this luminosity in the central $\sim 2$ pc of R136 \citep{Massey1998}, but this still corresponds to a mean interstellar separation $> 10^4$ AU. Thus the forces deposited in the small regions where our approximation $\kappa(T)\propto T^2$ does not apply will not tend to be dynamically dominant, since the forces exerted by different stars will tend to cancel one another.

\section{Results for convective versus radiative stability limits}
\label{sec:RTappendix}

\begin{figure}%[hb!]
\centering
\includegraphics[width = 0.5 \textwidth]{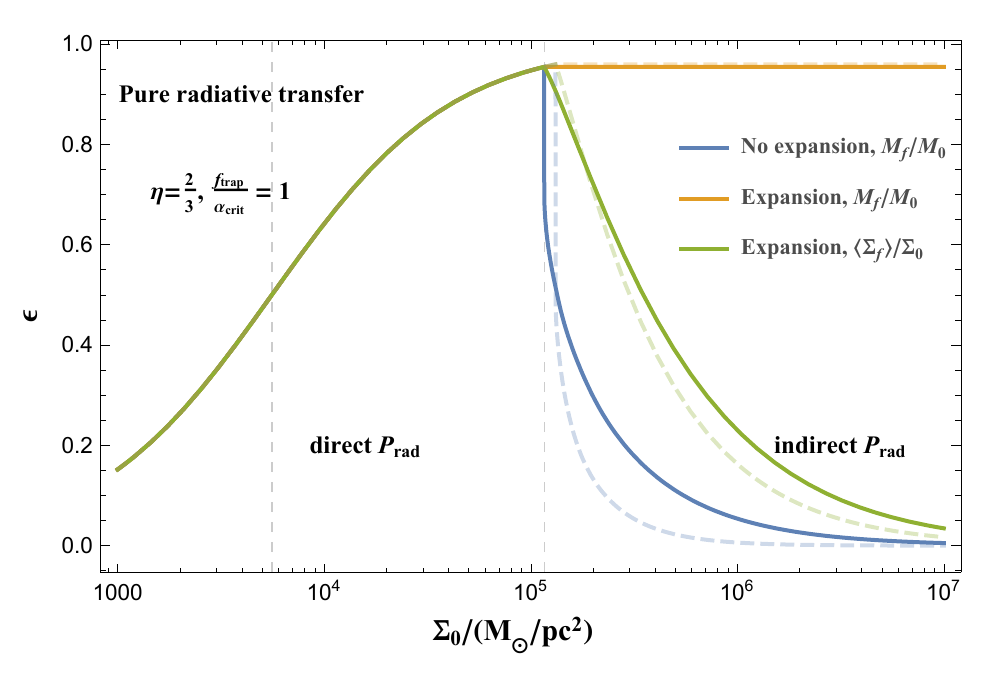}
\caption{Final star-formation efficiency for sub-regions as a function of initial mass surface density, $\Sigma_0$, for the limiting assumptions of efficient versus inefficient convection. This figure is identical to \autoref{fig_plotFinalEfficiencyDisplay} of the main text, and is calculated identically, except that the solid lines use the stability limit $f_{\rm E,crit} = f_{\rm E,crit,rt}$ (inefficient convection) rather than $f_{\rm E,crit} = f_{\rm E,crit,c}$ (efficient convection). For comparison, the faint dashed lines show the efficient convection case shown in \autoref{fig_plotFinalEfficiencyDisplay} of the main text.
} 
\label{fig_plotFinalEfficiencyRTDisplayRT}
\end{figure}

As described in the main text, \citetalias{Crocker2018}'s calculation of the critical surface Eddington ratio $f_{\rm E,crit}$ at which systems transition from stable and sub-Eddington to unstable and super-Eddington depends weakly on the assumptions one makes about the effectiveness of convective heat transport in fluid where radiation pressure dominates the enthalpy budget, and thus conventional fluid convection is ineffective. All figures shown in the main text, except where noted, are for the ``efficient convection" where one assumes that, despite radiation dominating the enthalpy budget, convection is nonetheless able to completely flatten the entropy gradient. To justify our assertion that the results would not change qualitatively under the opposite assumption that convection is unable to alter the entropy gradient in this regime, in \autoref{fig_plotFinalEfficiencyRTDisplayRT} we repeat our calculation of the final star formation efficiency using this alternative assumption. In this figure, the solid lines show the case assuming inefficient convection, while the dashed lines show the results assuming efficient convection as in the main text. Clearly there is little qualitative difference between the results in the two cases.

\section{The effects of a clumpy gas distribution}
\label{app:clumpy}

In \citetalias{Crocker2018}, we did not explicitly consider the effects of gas clumping on the column density, because we could simply envision our calculation as applying locally to every patch of a galactic disc. In spherical geometry, however, the column is necessarily an averaged quantity, and thus we must consider how inhomogeneities in a cloud are likely to affect the gas column density $\Sigma$.

Up to the point where the radiation flux becomes large enough to start ejecting gas, variations in $\Sigma$ will be driven primarily by turbulence. Thus for the purposes of calculating at what point radiation is capable of ejecting mass, the column density distribution that is relevant is the one induced by turbulence, which \citet{Thompson2016} show is well-described by a probability distribution function (PDF) that is approximately lognormal, with a dispersion given by
\begin{equation}
\sigma_{\ln\Sigma}^2 \approx \ln \left(1 + R \mathcal{M}^2/4\right).
\end{equation}
Here $\mathcal{M}$ is the Mach number of the turbulence,
\begin{equation}
R = \frac{1}{2}\left(\frac{3-\alpha}{2-\alpha}\right) \left[\frac{1-\mathcal{M}^{2(2-\alpha)}}{1-\mathcal{M}^{2(3-\alpha)}}\right],
\end{equation}
and $\alpha\approx 2.5$ is the index of the density power spectrum. For such a distribution, the area-weighted median column density is lower than the area-weighted mean by a factor of $\exp\left(-\sigma^2_{\ln\Sigma}/2\right)$, while the mass-weighted median is higher by the same factor. For $\mathcal{M} < 100$, this factor is $<3.7$, i.e., the median column density is within a factor of 4 of the mean column density. Thus the variations in column density induced by turbulence, as opposed to the variations in volume density, are relatively modest, and comparable to our uncertainties in other quantities (e.g., the dust opacity). Our estimated final column densities should be regarded as uncertain by roughly this factor.

There remains the possibility that these modest variations in column density will have substantially larger effects on the interaction of radiation with the gas, due to preferential escape of photons through low-density regions. However, radiation-hydrodynamic simulations show that this effect is surprisingly small, and does not substantially alter the conditions for gas to be ejected compared to the results of a naive laminar calculation \citep{Davis2014, Tsang2015, Zhang2017}.

\end{appendices}

% Don't change these lines
\bsp	% typesetting comment
\label{lastpage}
\end{document}